# THEORETICAL PREDICTION AND EXPERIMENTAL MEASUREMENT OF THE MIXED FLOCCULATION/COALESCENCE RATE OF IONIC HEXADECANE-IN-WATER NANO-EMULSIONS

German Urbina-Villalba*, Neyda García-Valera, Kareem Rahn-Chique

Instituto Venezolano de Investigaciones Científicas (IVIC), Centro de Estudios Interdisciplinarios de la Física, Lab. de Fisicoquímica de Coloides, Apartado 20632, Edo. Miranda, Venezuela.

**Abstract**     Theoretical calculations of the mixed aggregation/coalescence ($k_{FC}$) rate corresponding to a set of hexadecane-in-water nano-emulsions stabilized with sodium dodecyl sulphate (SDS) at different NaCl concentrations are presented. The rates were obtained through the change of the total number of aggregates of the dispersions as a function of time, predicted by Emulsion Stability Simulation (ESS). Two different models were implemented in order to mimic the dependence of the surface excess of the surfactant on the salt concentration. Experimental measurements of $k_{FC}$ were also made, based on the change of the turbidity of the emulsions as a function of time. A satisfactory agreement between theory and experiment is only attained if the model of surfactant adsorption accounts for the balance between the salting out of the surfactant solution and the partial screening of the surface charge of the drops induced by the increase of the ionic strength of the continuous phase. The observed behavior cannot be justified on the grounds of the Derjaguin-Landau-Verwey-Overbeek (DLVO) theory. Instead, the reversible flocculation of the aggregates of any size is proposed as an alternative mechanism to explain the dependence of $k_{FC}$ as a function of the salt concentration.

## 1. INTRODUCTION

Liquid/liquid dispersions (emulsions) are ubiquitous in our daily life. They are unstable systems which separate into two phases. These phases are generically identified as oil and water in order to emphasize the characteristic immiscibility between an organic and a polar liquid. Emulsions constitute food, cosmetics, medicines, and are present in both products and supplies of most industries. Some of them are generated on purpose, and some others occur naturally due to the generalized use of water for the cleaning and the disposal of organic waste.

Although most non-biological mechanisms of destabilization of oil-in-water (O/W) emulsions are well known, the prediction of their long time behavior is difficult. Such uncertainty is partially due to the fact that all destabilization processes are connected, occur simultaneously, and their relative importance changes as a function of time. Additionally, it is also owed to our inability to estimate the rates of incidence of each of these processes. This is particularly relevant in the case of drops of nano-metric size, which are impossible to be observed by optical methods. Hence, their aggregation, coalescence, or growth by molecular exchange cannot be directly detected, until the drops reach sizes of the order of a micrometer. Furthermore, the estimation of the ripening rate is based on the augment of the average radius of the drops as a function of time. However, that increase is also the product of the coalescence of the drops. Moreover, most of the equipments available only estimate the average hydrodynamic radius of the objects suspended in the liquid, and cannot discriminate between an aggregate of drops of a given hydrodynamic radius, and one big drop of the same radius. Hence, the rate of change of the average radius of an emulsion may be either related to flocculation, coalescence or Ostwald ripening, and therefore, it does not constitute a reliable measurement of any specific process. It is however, the best measurement of the overall stability of the system, because it separates the effect of gravity from the effect of other mechanisms of destabilization [Mendoza, 2013].

In order to understand the behavior of liquid dispersions, Emulsion Stability Simulations (ESS) can be a very valuable tool. The first calculation goes back to the year 2000 [Urbina-Villalba, 2000a]. In this seminal publication we showed that if the drops coalesce immediately after flocculation, the variation of the total number of drops as a function of time follows the equation of Smoluchowski [Smoluchowski, 1917]. Following that report, several articles with generic simulations regarding theoretical aspects of the emulsion stability problem were published. Among other topics, the effect of: inhomogeneous surfactant distributions [Urbina-Villalba, 2000b, 2001a, 2001b], polydispersity and thermal exchange [Urbina-Villalba, 2003a], volume fraction of internal phase [Urbina-Villalba, 2004a], time-dependent surfactant adsorption [Urbina-Villalba, 2004b], hydrodynamic interactions and effective diffusion constants [Urbina-Villalba, 2003b], and steric potentials [Lozsán 2005, 2006], were addressed. It became evident, that:

a) As an emulsion coalesces its total interfacial area diminishes. Since the surfactant concentration in a system is constant, this means that the bulk surfactant concentration progressively increases. Hence, the adsorption equilibrium of the surfactant has to be modified, inducing a progressive increase on the surface charge of the drops. This mechanism would lead to equilibrium in the absence of other destabilization processes. However, the increase of the average radius of the drops promotes creaming, deformation upon flocculation, and the occurrence of coalescence through additional mechanisms, like capillary waves and the appearance of holes [Vrij, 1968; Kashchiev, 1980; Osorio, 2011].

b) It is unclear whether the mechanisms of capillary waves or hole formation occur during the coalescence of non-deformable droplets, but up to our knowledge there is no experimental evidence of such phenomenon. Hence, it is assumed that spherical non-deformable drops coalesce when they flocculate in the primary minimum, since there are no additional repulsive barriers to prevent the drainage of the remaining liquid film between aggregated drops. Unlike deformable drops, non-deformable drops do not need to change their curvature in order to provide a path to connect their inner liquids. Therefore, an additional energy for the coalescence of drops is not expected [de Vries, 1958].

c) The redistribution of surfactant molecules as a consequence of coalescence promotes the temporal evolution of the fractal coefficient of the aggregates formed [Urbina-Villalba, 2006a].

d) Large particles (with radius $R_i > 1$ microns) develop deep secondary minimum at high salt concentrations. This induces fast flocculation rates, comparable to the ones occurring in the absence of a repulsive barrier [Urbina-Villalba, 2005b].

e) The deformability of the drops depends on many variables. However, an approximate threshold for drop deformation corresponds to a radius of 2.5 microns [Rojas, 2010a, 2010b, Toro-Mendoza, 2010]. Particles of lower radius tend to behave as rigid spherical particles.

f) The condition of stability of lyophobic colloids, originally enunciated in terms of the Derjaguin-Landau-Verwey-Overbeek (DLVO) theory [Derjaguin, 1941; Verwey, 1948] as: $V_T = 0$, $\partial V_T / \partial r_{ij} = 0$ (where $V_T$ is the total interaction potential between particles i and j, equal to the sum of attractive and electrostatic contributions $V_T = V_A + V_E$), has to be modified in order to take into account those maxima of the potential which may occur at $V_T < 0$. Such maxima also stabilize the system towards irreversible aggregation. Hence, the condition of stability should be substituted by: $\partial V_T / \partial r_{ij} = 0$ with $\partial^2 V_T / \partial r_{ij}^2 < 0$, that is, by the requirement of the existence of a maximum value [Urbina-Villalba, 2009a].

g) The fact that in a Brownian movement the energy given by the solvent to move a particle is removed from the particle as it diffuses through the liquid, causes the particle to move close to their potential energy curve at all times.

h) Large random deviates corresponding to the outskirts of the Gaussian distribution of the Brownian movement of relatively large particles (in comparison to the size of the solvent molecules) are necessary in order to reproduce the experimental value of the diffusion constant of Stokes. These deviates are equivalent to huge energy fluctuations in terms of the thermal energy $k_B T$ (where $k_B$ is the Boltzmann constant and T the absolute temperature) [Urbina-Villalba, 2003a].

i) Large repulsive barriers can be surpassed by the effect of the random kicks. The threshold for this process depends on the slope of the potential energy curve and the width of the repulsive barrier. For this mechanism to be effective, the particle has to reach close distances of separation. This means "crawling" on top of the potential energy curve, and then "jumping" over the maximum pushed by a thermal fluctuation.

j) When non-deformable droplets aggregate in the primary minimum and coalesce, it is not possible to identify

when the process of flocculation finishes and the process of coalescence begins. Hence, an evaluation of the aggregation rate includes the process of coalescence.

k) For typical diffusion coefficients ($10^{-10}$ m²/s), a surfactant concentration of 5 x $10^{-4}$ M is enough to build a considerable repulsive barrier between the drops in a fraction of a second [Urbina-Villalba, 2004b]. This is believed to be related to the previous findings of Rosen in regard to stability [Rosen, 1990].

Unlike older papers, our most recent publications seek the direct comparison of the simulations with experimental measurements in order to validate the relevance of the ESS method for studying the behavior of colloidal systems. In Ref. [Urbina-Villalba, 2009b] the stability ratio of a suspension of 96-nm anionic latex particles (at several salt concentrations) was reproduced, using both two-particle and many-particle simulations. In Ref. [Urbina-Villalba, 2009c] the process of Ostwald was included in the code, in order to study the short time evolution of a dodecane in water emulsion in the absence of surfactants. Among other interesting results, those computations showed that it is possible to obtain a linear variation of the cube average radius of an emulsion as a result of flocculation and coalescence. Moreover, it was found that the average radius fluctuates up and down an average slope [Nazarzadeh, 2013]. This is the outcome of two opposite phenomena: a) the molecular exchange between the drops which induces a *decrease* of the average particle radius at all times (unlike predicted by the LSW theory [Lifshitz, 1961; Wagner, 1961]), and b) the elimination of particles either by complete dissolution or coalescence that increases the average radius of the dispersion.

In Ref. [Osorio, 2011] we revisited a classical paper regarding the effect of droplet deformability on the stability of (decane + CCl$_4$)/water emulsions stabilized with either sodium oleate or sodium di-octil sulfosucciante. It was shown that the differences in the values of the flocculation rates of these two systems cannot be ascribed to differences in the deformability of the drops as formerly claimed. In Refs. [Rojas, 2010a, 2010b] ESS were used to reproduce the average coalescence time of nondeformable drops of hexadecane stabilized with β-casein (1 – 10 microns) with its planar homophase. Similarly, our most advanced simulations successfully reproduced the average lifetime of large deformable drops (1 – 1000 microns) of soybean oil stabilized with Bovine Serum Albumin with a planar interface. In those simulations, the deformability of the drops changed with time during their approach to the planar interface as a product of a time-dependent surfactant adsorption. Strong evidence in favor of the coalescence mechanism proposed by Gosh and Juvekar was provided [Gosh, 2002].

During the last years our research had concentrated in the development of a novel procedure for the determination of an average flocculation/coalescence rate ($k_{FC}$) of nanoemulsions using turbidity measurements. In the typical case, the effect of ripening is minimum during the time of measurement (typically 60 seconds), the value of $k_{FC}$ essentially correspond to the processes of flocculation and coalescence. According to the methodology proposed [Rahn-Chique, 2012], a value of $k_{FC}$ is obtained minimizing the difference between the experimental variation of the turbidity of the emulsion as a function of time, and a theoretical expression of the turbidity which accounts for the dispersion of light by aggregates of primary drops and separate (disaggregated) spherical drops of each possible size:

$$\tau(t) = n_1(k_{FC}, t)\sigma_1 + \sum_{k=2}^{k_{max}} n_k(k_{FC}, t)[x_a \sigma_{k,a} + (1-x_a)\sigma_{k,s}]$$

(1)

Here $n_k$ stands for number of aggregates of size k predicted by Smoluchowki [Somoluchowski, 1917] for the case of irreversible aggregation:

$$n_k = \frac{n_0 (k_{FC} n_0 t)^{k-1}}{(1 + k_{FC} n_0 t)^{k+1}}$$

(2)

Parameter "$x_a$" represents the fraction of (non-spherical) aggregates of primary drops existing at each time.

The method outlined above is now commonly used in our laboratory, and allows evaluating the variation of the $k_{FC}$ –corresponding to a set of hexadecane-in-water dispersions stabilized with sodium dodecyl sulphate (SDS)– as a function of the ionic strength. In this paper we test the soundness of Emulsion Stability Simulations for reproducing that experimental data. As will be evident from the results, the agreement is fair as long as an appropriate algorithm for the distribution of surfactant molecules is used [Urbina-Villalba, 2013]. When this is accomplished, the simulations reasonably agree with the experimental measurements, providing additional information on the type of the aggregates formed, the change of the average radius, the total interfacial area, the degree of ripening, and several other relevant variables whose knowledge might be very valuable in order to understand the long-time behavior of these systems.

## 2. THEORETICAL BACKGROUND: THE ALGORITHM OF EMULSION STABILITY SIMULATIONS

For a detailed description of Emulsion Stability Simulations the reader is referred to the following articles [Urbina-Villalba, 2000, 2003b, 2004c, 2009a; Toro-Mendoza, 2010; Osorio, 2011].

ESS are based on the algorithm of Ermak and McCammon [Ermak, 1978] for Brownian Dynamics simulations. However, the implementation of routines suitable for: a) the calculation of the effective diffusion constants with account of hydrodynamic interactions between the drops, b) the handling of the surfactant population, c) the preservation of a minimum number of drops during the calculation, e) the incorporation of the steric potential between non ionic surfactant chains, and f) the inclusion of the processes of drop deformation, coalescence, Ostwald ripening, capillary waves, etc, produced a substantially different computational technique that we now recognize with the name of Emulsion Stability Simulations. Hence, ESS identifies the way in which we do our simulations regardless of any restriction imposed by the original algorithm of Brownian dynamics.

As in Brownian Dynamics, the displacement of drop "i" during time $\Delta t$: $\vec{r}_i(t + \Delta t) - \vec{r}_i(t)$, results from the effect of :

1. Deterministic inter-particle forces $\sum_{i=1; j \neq i}^{N} \vec{F}_{ji}$. These include the van der Waals attraction between the molecules of oil composing the drops, and the repulsive interaction between the surfactant molecules adsorbed to the interface of the drops. It also, comprehend the application of external fields as in the case of gravity, $\vec{F}_{ext}$.

2. Random forces resulting from the thermal exchange between the drops and the surrounding liquid media (implicitly considered). These are the ones that produce the Brownian movement. The statistical properties of this movement are reproduced generating a vector of random numbers which belongs to a Gaussian distribution of real numbers with zero average and unit variance. The typical square mean displacement is obtained, multiplying each deviation by $\sqrt{2\, D_{eff, i}\,(d_c\,,\,\phi)\Delta t}$. Where $D_{eff\!\!\!f}\,(d_c\,,\,\phi)$ is an effective diffusion constant which takes into account the hydrodynamic interaction between drops, and is a function of a characteristic distance of approach ($d_c$) and the volume fraction of oil in the immediate neighborhood of each drop ($\phi$):

$$D_{eff, i}\,(d_c\,,\,\phi) \;=\; D_0\, f_{corr, i} \;=\; (k_B\,T\,/\,6\,\pi\,\eta\,R_i)\, f_{corr, i} \tag{3}$$

Here: $R_i$ is the radius of drop i, $\eta$ is the viscosity of the external phase (water in the present case), $k_B$ is the Boltzmann constant, $T$ the absolute temperature. $f_{corr, i}$ is a hydrodynamic (mean field) correction [Urbina-Villalba, 2003b]. If a drop "j" gets nearer than $r_{ij} - R_i - R_j \leq R_i$ Honig's et al. [Honig, 1971] formula is used to calculate the diffusion constant of the drop i:

$$f_{corr, i} \;=\; (6\,u^2 + 4\,u)\,/\,(6\,u^2 + 13\,u + 2) \tag{4}$$

Where: $u = (r_{ij} - R_i - R_j)\,/\,R_R$ and,

$$R_R = 1/2\,(R_i + R_j) \quad \text{otherwise:}$$

$$f_{corr, i} \;=\; 1.0 - 1.734\,\phi + 0.91\,\phi^2 \tag{5}$$

In the former implementations of the program Eq. (5) was evaluated using the local volume fraction of oil around each particle. In the present version, the volume fraction of oil in the whole simulation cell is used for this purpose.

Using an effective diffusion constant to modulate the velocity of the particles, the equation of motion of drop "i" can simply be expressed as:

$$\vec{r}_i(t + \Delta t) \;=\; \vec{r}_i(t)$$

$$+ \left\{ \left( \sum_{j=1, j \neq i}^{N} \vec{F}_{ji} + \vec{F}_{ext} \right) (D_{eff, i}\,(d_c\,,\,\phi)\,/\,k_B\,T) \right\} \Delta t$$

$$+ \sqrt{2\,D_{eff, i}\,(d_c\,,\,\phi)\,\Delta t}\,\;[\vec{G}auss] \tag{6}$$

Here the term in the parenthesis on the right hand side has dimensions of velocity and stands for the average velocity experienced by the particle during a time $\Delta t$. The displacement of a particle: $\Delta r = \left| \vec{r}_i(t + \Delta t) - \vec{r}_i(t) \right|$ during this time is the product of the deterministic forces and a random "kick" produced by the (implicit) collisions of the surrounding solvent molecules with the surface of the particle.

The interaction forces of Eq. (6) are obtained deriving the potentials of interaction as a function of the intermolecular distance $r_{ij} = \left| \vec{r}_i(t) - \vec{r}_j(t) \right|$:

$$F_{ij} = -dV(r_{ij}) / dr_{ij} \qquad (7)$$

The attractive interaction between the drops of oils is modeled using the expression of Hamaker [Hamaker, 1937] for two spherical drops:

$$V_A = V_{vdW} =$$
$$- A_H / 12 \left( y / (x^2 + xy + x) + y / (x^2 + xy + x + y) \right.$$
$$\left. + 2 \ln \left[ (x^2 + xy + x) / (x^2 + xy + x + y) \right] \right) \qquad (8)$$

Here: $x = h/2R_i$, $y = R_i/R_j$, and $A_H$ is the Hamaker constant.

In the case of an ionic surfactant like SDS, the electrostatic potential at the surface of each drop $\Psi_0$ can be calculated from the surface charge density: $\sigma_{elec}$. The expressions that connect these two variables are generally involved, and result from the evaluation of the electrostatic potential or its derivative (resulting from a particular solution of the Poisson-Boltzmann equation) at the surface of the particle. An approximate relationship between the electrostatic potential and the surface charge density was provided by Sader [Sader, 1997]:

$$\sigma_{elec} e / \kappa \varepsilon \varepsilon_0 k_B T =$$
$$\Phi_p + \Phi_p / \kappa R_i - \kappa R_i \left( 2 \sinh (\Phi_p / 2) - \Phi_p \right)^2 / \overline{Q} \qquad (9)$$

In the expression above $\varepsilon_0$ is the permittivity of vacuum, $\varepsilon$ the dielectric constant of water, $\kappa$ the inverse of the Debye length and:

$$\overline{Q} = 4 \tanh (\Phi_p / 4) - \Phi_p - \kappa R_i \left[ 2 \sinh (\Phi_p / 4) - \Phi_p \right] \qquad (10)$$

$$\Phi_p = \Psi_0 e / k_B T \qquad (11)$$

The evaluation of Eq. (9) is the most time consuming step of the simulation. The bisection method is employed to calculate the electrostatic surface potential of each particle in terms of its surface charge density. In the general case, the surface charge changes with time due to the variation of the total interfacial area of the emulsion and/or the use of the routine of time-dependent surfactant adsorption.

Equation (9) is valid at relatively long distances (far field) but it's applicable to both high and low surface potentials. The electrostatic free energy at each separation can be calculated from [Danov, 1993]:

$$V_E = (64 \pi C_{el} k_B T / \kappa) \tanh (e \Psi_{0i} / 4 k_B T)$$
$$\tanh (e \Psi_{0j} / 4 k_B T) \quad x$$
$$exp(-\kappa h) \left[ 2 R_i R_j / \kappa (R_i + R_j) \right] \qquad (12)$$

In the absence of surfactants the surface potential results from the preferential adsorption of hydroxyl ions to the surface of the drops [Beattie, 2004; Stachurski, 1996; Marinova, 1996].

Equations (3) – (12) provide the information necessary for the movement of the particles (Eq. (6)). Whether two particles flocculate or not depends on the characteristics of their total interaction potential: $V_T = V_E + V_A$ (Eqs. (8) and (12)). In the case of non deformable droplets, coalescence occurs if the distance between the center of the drops becomes smaller than the sum of their radius: $r_{ij} \leq R_i + R_j$. Thus, a resulting particle is formed at the center of mass of the former coalescing drops, which are then removed from the simulation box. The radius of the new drop is calculated from the preservation of the volume of the former drops ($R_{new} = \sqrt[3]{R_i^3 + R_j^3}$). Three dimensional periodic boundary conditions are used in all calculations. Thus, a drop can flocculate and/or coalesce with either "real" particles inside the simulation box, or their periodic boundary images.

Prior to the coalescence check, the Ostwald ripening routine allows the particles to exchange (gain or lose)

molecules of oil depending on the relation between their actual radius and the average radius of the emulsion at a particular time.

Our implementation of the Ostwald ripening process uses the algorithm of De Smet [DeSmet, 1997] for this process. However, unlike De Smet's implementation, the number of transferred molecules at each step depends on the time step of the simulation and it is not arbitrary.

Following Fick´s law, using the equation of Kelvin, and assuming that the capillary length $\alpha$ is substantially smaller than the particle radius [De Smet, 1997], it can be deduced that the number of molecules (m) of oil in drop i: $m_i$, changes with time according to:

$$\frac{dm_i}{dt} = 4\pi D_m C_\infty \alpha \left( \frac{R_i}{R_c} - 1 \right) \qquad (13)$$

Where:

$$\alpha = 2\gamma V_{molar} / RT \qquad (14)$$

is the capillary length, $\gamma$ is the interfacial tension of the drops, $V_{molar}$ is the molar volume of oil, $C_\infty$ is the solubility of the oil in the presence of a planar interface, $D_m$ is the diffusion constant of the oil in the water phase, and $R_c$ is the critical radius of the emulsion, which is equal to the number average radius of the dispersion.

During the simulation the average radius ($R_c$) changes as a function of coalescence and Ostwald ripening. A drop will decrease or augment its radius $R_i$ at a particular time, depending on whether $R_i$ is smaller or larger than $R_c$ (Eq. (13)).

Since the number of drops decreases when coalescence occurs, the coupling of the algorithm of Ostwald proposed by De Smet with our former ESS code, required a new routine for preserving a minimum number of particles during the calculation. This is not necessary if the conditions are such, that the number of particles is preserved (large repulsive potentials, low solubility of the oil, low surface tension, etc). Otherwise, when the initial number of particles ($N_0$) decreases to $N_0/4$, the simulation box is moved to the negative quadrant of the coordinate axis, and it is replicated three times using periodic boundary conditions [Urbina-Villalba, 2009a, 2009c]. The result is a bigger simulation box with $N_0$ drops. This new "macro" box exhibits the same particle size distribution occurring prior to its construction, while keeping constant the volume fraction of oil.

## 3. EXPERIMENTAL SECTION

The oil phase of the emulsion consisted on a mixture of hexadecane ($C_{16}H_{34}$, $\phi = 0.73$) with tetrachloroethylene (TCE, $\phi = 0.27$). This blend was prepared in order to approximate the density of the surrounding aqueous solution ($d_{mix} = 0.988$ /cm$^3$), thus minimizing the effect of buoyancy during the measurements. An oil-in-water nanoemulsion ($R_i = 184$ nm, C.V.= 22%) was produced using the method of phase inversion by a composition change [Solé, 2006]. The starting "concentrated" system contained 0.136 M of NaCl, 1.2% of sodium dodecyl sulphate, and a volume fraction of $\phi = 0.093$. These conditions correspond to a two-phase system composed of a microemulsion and liquid crystal phase. The actual emulsion was obtained by dilution and slow agitation, adjusting the final SDS concentration to 0.5 mM, and the volume fraction to $\phi = 9$ x $10^{-4}$ ($n_0 = 3.6$ x $10^{16}$ m$^{-3}$). The drop size distribution obtained was lognormal with an average radius of 184 nm. In the absence of additional salt (see below), the drops exhibited a zeta potential of $-78.6$ mV.

The phenomenon of flocculation was studied taking 2.6 ml aliquots of the dilute emulsion and adjusting its salt concentration to a specific value between 300 and 1000 mM of NaCl. Prior to this, a small amount of salt was added in order to fix the initial ionic strength of the solution to a fixed value of 10.5 mM. In other words: $I_s = 10.5$ mM $= C_s + C_{NaCl}$ (where $I_s$ stands for the total ionic strength, $C_s$ correspond to the surfactant concentration, and $C_{NaCl}$ to the concentration of sodium chloride). The variation of the turbidity as a function of time was measured using an UV-Visible spectrophotometer (UV-1800, Shimatzu) at $\lambda = 800$ nm. This wavelength was selected in order to minimize the degree of light adsorption by the oil, and fulfill the conditions for the applications of Raleigh-Gans-Debye theory of light scattering [Rahn-Chique, 2012]. The data was adjusted to Eq. (1) using a package of symbolic algebra (Mathematica, 8.0.1.0). This fitting is not trivial due to the complex form of the optical cross sections of the aggregates, and their dependence on several parameters, including the radius of the drops, the index of refraction of the particles and the one of the surrounding medium, and the wavelength of light employed. The measurements were repeated at least three times.

## 4. COMPUTATIONAL DETAILS

The parameters of the simulations are shown in Table 1. The value of the Hamaker constant for hexadecane was used. However it is expected to increase due to the presence of TCE in the oil mixture. The solubility of the oil is also expected to be larger than the one of pure hexadecane. An extremely small time step was necessary in order to sample the interaction potential appropriately. In order to select the time step several preliminary 5-second runs were made. The value of $\Delta t$ selected corresponded to the maximum value of the time step for which a plot of $R_{ave}(t)$ vs. t, did not show appreciable variations with a further decrease of $\Delta t$. In that case, a denser sampling of the interaction potential using a slower time step was believed to be unnecessary.

Table 1: Parameters common to **Set I**
and **Set II** Simulations

| Parameter | Value |
|---|---|
| $A_H$ (J) | $5.40 \times 10^{-21}$ |
| $C_\infty$, cm³/ cm³ | $2.72 \times 10^{-11}$ |
| $D_m$ (m²/s) | $4.60 \times 10^{-10}$ |
| $V_m$ (m³/mol) | $2.92 \times 10^{-4}$ |
| q (e⁻) | -0.09 (**Set I**) |
| | -0.26 (**Set II**) |
| Cut off length (µm) | 5.0 |
| r (g/cm³) | 0.775 |
| $\Delta t$ (s) | $7.94 \times 10^{-8}$ |
| $N_0$ | 500 |
| $\phi$ | $9.5 \times 10^{-4}$ |

In order to study the behavior of the nanoemulsions, a SDS concentration of $C_s = 0.5$ mM was selected (CMC = 8.3 mM). The ionic strength was adjusted to 10.5 mM with NaCl in order to allow the comparison of the present results with those of latex particles with similar surface charges. Hence, the actual value of the ionic strength in the simulations and the experiments is the sum of a pre-selected target concentration (300 to 1000 mM NaCl) plus 10.5 mM (0.5 mM SDS + 10.0 mM NaCl).

Two sets of simulations were run. The first set (**Set I**) uses a routine which distributes the surfactant molecules on the surface of the drops according to the surface excess suggested by macroscopic adsorption isotherms. For this purpose the empirical equation of the interfacial tension proposed by Gurkov *et al.* [Gurkov, 2005] was used. This expression is a generalization of the one previously deduced by Rehfeld [Rehfeld, 1967] from experimental measure-

ments of the interfacial tension of SDS at a hexadecane/water interface:

$$\gamma = 0.0401 \left(\ln(a_s \, a_t)\right)^3 + 0.7174 \left(\ln(a_s \, a_t)\right)^2$$
$$- 6.9933 \left(\ln(a_s \, a_t)\right) - 89.1414 \tag{15}$$

Here:

$$a_s = \gamma_\pm \, C_s \tag{16}$$

$$a_t = \gamma_\pm \, (C_s + C_{NaCl}) \tag{17}$$

$$\gamma_\pm = \frac{A \sqrt{I_s}}{1 + B d_t \sqrt{I_s}} + b I_s \tag{18}$$

In Eqs. (15) – (18): $I_s$ stands for the ionic strength (in M); A = 0.5115 M⁻¹/², $Bd_t$ = 1.316 M⁻¹/², b = 0.055 M⁻¹. The value of the surface excess results from differentiation of Eq. (15) with respect to $\ln(a_s a_t)$:

$$\Gamma = \frac{0.001}{k_B T} \left( 0.1203 \left(\ln(a_s \, a_t)\right)^2 \right.$$
$$\left. + 1.4348 \left(\ln(a_s \, a_t)\right) - 6.9933 \right) \tag{19}$$

To complete the isotherm, the critical micelle concentration was approximated using Corrin-Harkins relationship [Corrin, 1947; Urbina-Villalba, 2013]:

$$Log_{10} CMC = -0.45774 \, Log_{10}(C_{NaCl}) - 3.2485 \tag{20}$$

The value of the tension at the CMC was approximated by the expression of Aveyard *et al.* [Aveyard, 1987]:

$$\gamma_c = -0.77969 \ln(C_s + C_{NaCl}) + 3.1458 \tag{21}$$

The number of surfactants at the interface of a drop is evaluated dividing the interfacial area of the drop ($A_i$) by $A_s = \Gamma^{-1}$ (where $A_s$ the area of a surfactant molecule at the interface):

$$N_i = A_i / A_s \tag{22}$$

In the event that the number of surfactants was not enough to cover the interface of the drops, then a fraction of the total number of surfactants ($N_s$) was assigned to each drop:

$$N_i = N_s \left( A_i \Big/ \sum A_i \right) \qquad (23)$$

Equations (15) to (19) are only valid in a certain range of concentrations and ionic strengths. The routine employed for the calculations of **Set I** makes specific provisions for this problem: If either $\gamma < 0$ or $\Gamma^{-1} = A_s < A_\infty$ (where $A_s$ the area of a surfactant molecule at the interface, and $A_\infty$ the minimum area occupied by this molecule in the saturation limit: 37.7 Å$^2$) then the values of these variables are arbitrarily fixed to: $\gamma = 1$ Nm, and $A_s = A_\infty$. Figure 1 shows the adsorption isotherms predicted using the corrections specified in this paragraph. A more elaborate procedure for building the isotherms is shown in the Appendix.

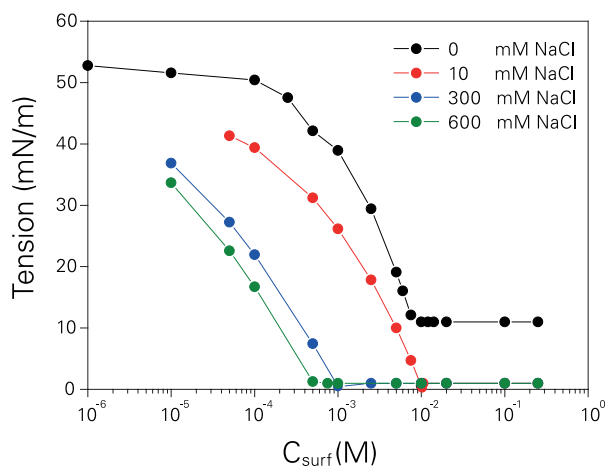

**Figure 1:** Adsorption Isotherms employed in the calculation of the surface excess for the simulations of **Set I**. They result from using Eqs. (15)-(23).

Once that the number of surfactants per drop can be calculated, the total charge of each drop ($Q_i$) can be estimated using:

$$Q_i = q_s e_0 N_i \qquad (24)$$

Where $e_0 = 1.6 \times 10^{-19}$ Coul, and $q_s$ is the *effective* charge of a surfactant molecule. Notice that the surface charge density of drop i (Eq. (9)) is simply: $\sigma_{elec} = \sigma_i = Q_i / A_i$. Hence, the surface potential of the drop can be evaluated solving Eq. (9): $\Psi_i = \Psi_i (Q_i)$. In a typical ESS calculation, the charge of the surfactant is found, reproducing the electrostatic po-tential of a drop at a fixed interfacial area. The program employs 13 different routines for apportioning the surfactant molecules to the interface of the drops. These are meant to replicate diverse experimental conditions.

The value of $q_s$ (Eq. (24)) used in the calculations of **Set I** ($q_s = -0.09$) was the one that reproduced the experimental surface potential of a 184-nm drop ($-78.6$ mV) belonging to a hexadecane-in-water emulsion with $C_s = 7.5$ mM and $I_s = 10.5$ mM.

The surface excess of the surfactant corresponding to the second set of calculations (**Set II**) is a byproduct of a routine implemented for reproducing the variation of the surface potential of hexadecane-in-water nanoemulsion drops as a function of SDS concentration and the ionic strength of the aqueous solution [Urbina-Villalba, 2013]. As theoretically predicted, it was observed that at a fixed surfactant concentration, the electrostatic surface potential shows a maximum as a function of the ionic strength of the solution (Figure 2). That maximum results from two opposite phenomena: the increase of surfactant adsorption promoted by the addition of salt (salting out), and the progressive screening of that surface charge. At low salt concentrations, the increase of the surface charge resulting from the additional surfactant adsorption induces a higher surface potential. However, as the surface gets saturated, only screening is possible, and the electrostatic potential necessarily decreases.

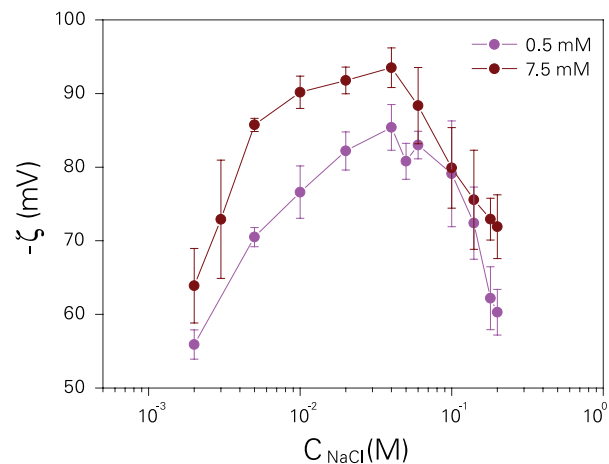

**Figure 2:** Change of the average zeta potential of the emulsions as a function of the salt concentration. The charge employed for the simulations of **Set II** is the one that reproduces the maximum value of the surface potential for $C_s = 7.5$ mM SDS.

If an adequate methodology is employed for reproducing the change of the surface potential as a function of the salt concentration, an empirical (approximate) equation for the calculation of the surface excess can be found:

$$\Gamma = G \ln C_{NaCl} + H \qquad (25)$$

The average values of G and H are: G = 6.72040 x $10^{-7}$ l/m², and H = 4.44776 x $10^{-6}$ l /m². The use of constant values for G and H is really an approximation [Urbina-Villalba, 2013], since these parameters are functions that exhibit a weak dependence on both $C_s$ and $C_{NaCl}$. The validity of Eq. (25) suggests a surfactant concentration equal or higher than 0.5 mM is enough to saturate the drops of a dilute $\phi = 10^{-4}$ emulsion. Hence, it is the salt concentration which regulates the surface excess. As a result, an "adsorption isotherm" in terms of the *salt concentration* results (Eq. (25)).

Despite the arguments of the previous paragraph, use of the Corrin-Harkins equation (Eq. (20)) indicates that the saturation of the interface may not occur under the present experimental conditions. The CMC changes from 1.60 to 0.56 mM SDS when the amount of salt increases from 100 to 1000 mM. Hence, in all the experiments reported, the surfactant concentration used lays below the CMC. Since surfactant aggregation generally occur after the maximum surface excess is attained, it is likely that the interface of the drops is not completely saturated.

Notice, that it is not possible to measure the zeta potential of an emulsion at ionic strengths higher than 200 mM NaCl using a Delsa SX-440. This is the general case. The methodology implemented allows extrapolating the value of the electrostatic surface potentials of the drops at ionic strengths that cannot be commonly reached experimentally [Urbina-Villalba, 2013]

The charge of the surfactant ($q_s = -0.26$) for the calculations of **Set II** was found, replicating the maximum value of the electrostatic potential (Figure 2) shown by a hexadecane-in-water emulsion containing 7.5 mM of SDS [Urbina-Villalba, 2013].

Figure 3 shows the potentials of interaction corresponding to the calculations of **Set I**. The repulsive barriers of the potentials are low, and diminish with the increase of the ionic strength as predicted by the DLVO theory. At the same time, the secondary minimum becomes progressively deeper, favoring the occurrence of fast aggregation rates.

Figures 4 to 5 show the interaction potentials corresponding to the calculations of **Set II**. The repulsive barriers of these potentials are considerably higher than the ones of **Set I**. The depth of the secondary minimum depends on the salt concentration, reaching values around $-10 \, k_B T$ for an ionic strength of 1 M. In fact, the repulsive barrier still surpasses 50 $k_B T$ for this extremely large concentration of salt. As will be shown, these barriers do not allow coagulation in the primary minimum, and therefore do not permit coalescence at least in the case of non-deformable droplets.

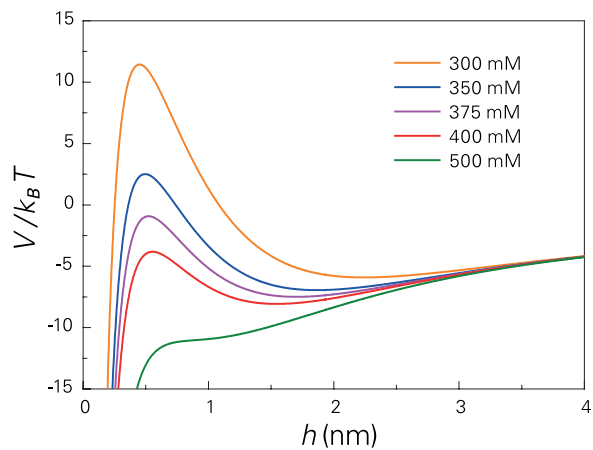

**Figure 3:** Interaction potentials predicted by experimental adsorption isotherms for the case of two 184nm-drops of hexadecane suspended in an aqueous solution of 0.5 mM SDS (**Set I**).

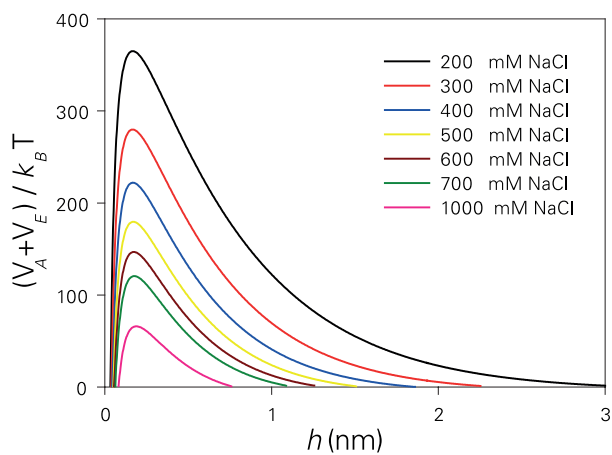

**Figure 4:** Interaction potentials deduced from zeta potential measurements for the case of two 184nm-drops of hexadecane suspended in an aqueous solution of 0.5 mM SDS (**Set II**).

Thus, the simulations parameterized using experimental data from macroscopic adsorption isotherms favor coalescence, while the ones parameterized using the variation of the zeta potential of the drops as a function of salt, do not.

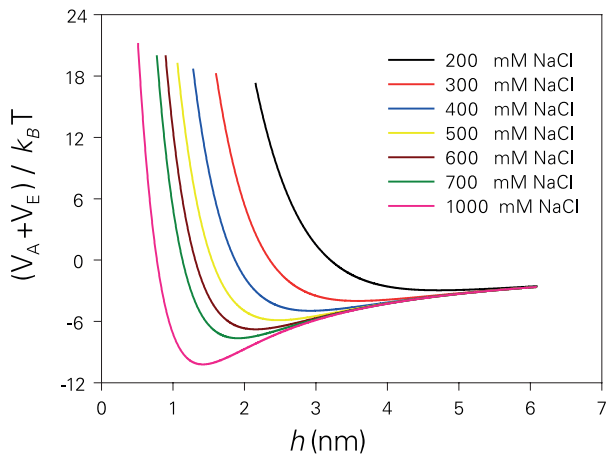

In order to compare the experimental values of $k_{FC}$ determined with the use of Eq. (1) with our simulations, the change of the number of aggregates of the emulsion as a function of time must be quantified. The simulations only provide the coordinates and the size of the drops existing at each time, but do not establish their degree of aggregation, nor do they keep track of their origin. However, the program records the variation of the number of particles as a function of time, and –in the event of complete particle dissolution– the number of particles eliminated by Ostwald ripening.

In the case of deformable drops, the change in the curvature of the drop identifies its degree of aggregation. However, the degree of aggregation of two spherical non-deformable drops (in the primary or secondary minimum) cannot be directly identified by inspection of the position of the drops at a given time. Two possible solutions of the problem are: a) to follow the variation in the relative position of the drops as a function of time, or b) to establish a specific distance of flocculation ($d_{floc}$) using a sound physical criteria, and then compare the actual distance between the drops with this threshold.

Careful observation of the interaction potential curves indicates that suitable values of $d_{floc}$ for **Set I** and **Set II** calculations are 3 nm and 4 nm, respectively. Notice, that if the flocculation distance is taken to be too long, some external particles will be mistakenly included in the floc. On the other hand, if the potential exhibits a larger repulsive barrier, $d_{floc}$ cannot be smaller than the position of the secondary minimum, since the particles will not be able to attain such distances, and no aggregates will be detected. Notice also that the position of the secondary minimum of the potential depends on the salt concentration. The selected values allow studying all the calculations of the same set with the same criteria. In any event, this is an approximate way to determine the existence of flocs, since the position of the secondary minimum changes as the particles grow in size. On the other hand, because the method is applied to all sets of coordinates stored, the fortuitous approach of two particles, as well as reversible flocculation will be observed by the temporal variation of the total number of aggregates.

The number of aggregates of each size was determined using a program previously developed by our group [Urbina-Villalba, 2005a]. The "coordinate file" contains the coordinates of the particles in the simulations stored periodically. The code reads the set of coordinates of the particles corresponding to each time (sequentially) and processes each block of coordinates separately. First, it identifies the number of neighbors that stand at a distance lower than $d_{floc}$ ($d_{ij} = r_{ij} - R_i - R_j < d_{floc}$) from the first particle. Then it repeats the same procedure for the particle's neighbors and its neighbors' neighbors until no other new member of the floc appears. At this point the number of particles composing the first aggregate had been identified and can be stored. Next, the program erases all the particles corresponding to this floc from the original set of coordinates, and then repeats the same (complete) procedure with the remaining particles.

Once the number of aggregates $n_{agg}$ is available, the theoretical value of $k_{FC}$ can be estimated from the average slope of $1/n_{agg}$ vs. t as predicted by Smoluchowski:

$$n_{agg}(t) = \frac{n_0}{1 + k_{FC}\, n_0\, t} \qquad (26)$$

The present calculations start from a Gaussian distribution of 500 drops randomly distributed inside a simulation box. The box is reconstructed every time the number of particles reaches 125 drops due to coalescence and/or ripening. This generates several sets of data corresponding to the variation of N from 500 to 125. In this event, Eq. (26) has to be applied to each set of data, and any changes in $k_{FC}$ as a function of time can be appraised. Instead, if the number of particles is preserved due to the existence of large potential barriers, the number of particles should remain equal to 500, and the evolution of the system can then be followed using the original cell, although –depending on the interaction potential– it is possible that at very long times, all the particles aggregate into one single cluster. This did not occur during the course of the present simulations. The total time of the simulation was 30 seconds. The experimental variation of the turbidity as a function of time extended for a period of 50 seconds.

The effect of gravity was accounted for in all simulations:

$$F_{ext} = F_{g'} = \frac{4}{3} \pi R_t^3 \Delta \rho \, g'$$ (27)

Here, $g'$ is the acceleration of gravity and $\Delta \rho$ the density difference between hexadecane and water. The effect of gravity is expected to be minor for nanometer drops during a lapse of the measurements. Since, the simulation box is of the order of a few microns, three dimensional periodic boundary conditions were used. This means that particles that escape the simulation box through the lid re-enter the box through its bottom, and vice versa. Hence, the simulation box resembles a slab of liquid in the middle of the actual recipient, and therefore, the calculations do not lead to the accumulation of drops at its upper end.

All calculations were run in a Dell Precision T7400 with 8 Xeon processors.

## 5. RESULTS AND DISCUSION

It is clear from Eq. (26) that the value of $k_{FC}$ results from the slope of $1/n_{agg}$ vs. time. In the absence of a strong repulsive barrier (**Set I** calculations) those correlations are very good ($r^2 > 0.99$) even when coalescence occurs. This is not surprising. In order to deduce Eqs (2) and (26), Smoluchowski assumed the absence of particles at the collision radius of a central particle. When the particles make contact, the program coalesces them instantaneously. Thus, as in the case of Smoluchowski, the collision radius is a perfect "sink".

It is important to remark that the code used for the classification of aggregates of different sizes, overestimates the number of monomers, since it identifies as a monomer any single particle that remains away from the rest, regardless of its radius. Thus, a drop that results from the coalescence of two monomers might be regarded as a singlet. However, in the experiment, it is counted as a dimer for the population balance of Eq. (1) since it contains twice the original volume of a monomer. Since the drop size distribution is polydisperse from the very beginning of the simulation, a more reliable code will need to keep track of the evolution of each particle as a function of time. This is very time consuming, especially when the drops are also allowed to change their radii through Oswald ripening.

Figure 6 illustrates the change in the number of singles, doublets, and triplets in the case of NaCl = 300 mM. In Figures 6 to 8, the ordinate axis corresponds to:

$$y_k = n_k(t) / n_0$$ (28)

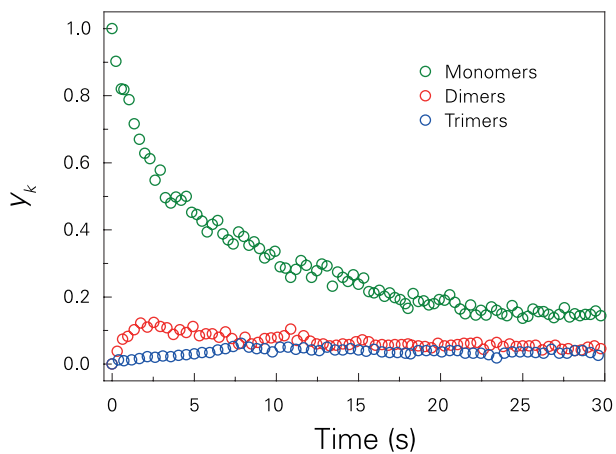

**Figure 6:** Change of the number of aggregates as a function of time predicted by the simulation of **Set I** corresponding 300 mM NaCl.

The figures referred above can be directly compared with the predictions of Smoluchowski for the case of rigid particles. Instead Table 2 lists the values of:

$$x_k = n_k(t) / n_{agg}(t)$$ (29)

**Table 2:** Fraction of aggregates of different size existing in the emulsion at t = 30 s.

| Set | NaCl (mM) | $x_1$ | $x_2$ | $x_3$ | $x_4$ | $x_5$ | $x_6$ |
|-----|-----------|-------|-------|-------|-------|-------|-------|
| I | 300 | 0.53 | 0.26 | 0.10 | 0.07 | 0.03 | 0.01 |
| II | 300 | 0.93 | 0.05 | 0.02 | 0 | 0 | 0 |
| II | 600 | 0.26 | 0.13 | 0.16 | 0.23 | 0.13 | 0.07 |

Equation (29) calculates the actual fraction of the aggregates at a given time. It is clear that $x_k$ is only equal to $y_k$ at t = 0 (n(t = 0) = $n_0$), because the total number of aggregates decreases as a function of time.

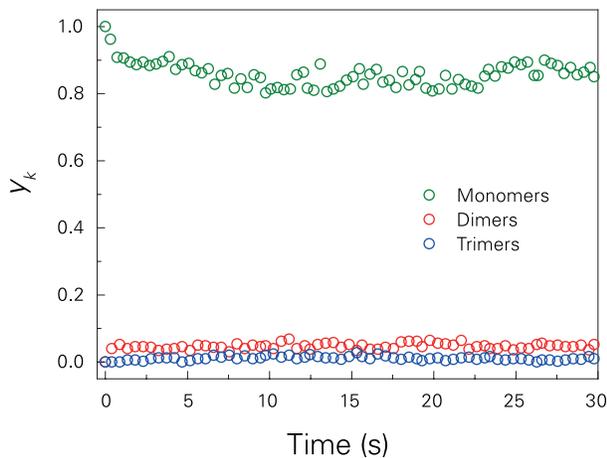

Figure 7: Change of the number of aggregates as a function of time predicted by the simulation of **Set II** corresponding 300 mM NaCl.

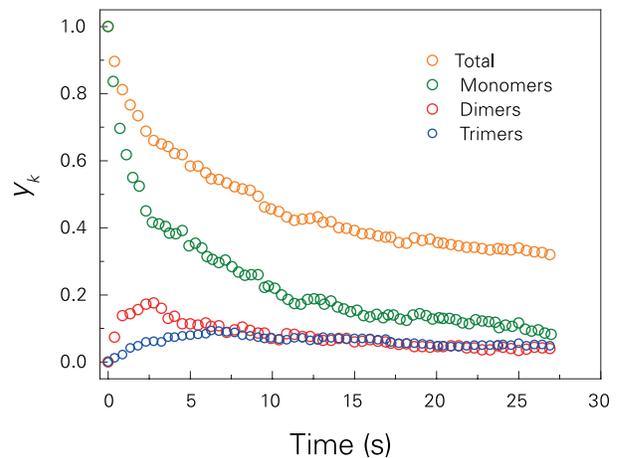

Figure 8: Change of the number of aggregates as a function of time predicted by the simulation of **Set II** corresponding 600 mM NaCl.

Despite the limitations produced by the small number of particles of the calculations (500) and those related to the classification of the flocs, the aggregates of each size follow the variation predicted by Smoluchowski (Eq. (2)). These calculations are faster than the ones of **Set II**, and therefore they were run for a total of 50 seconds, the typical lapse of time used in the experimental evaluation of $k_{FC}$.

Unlike the regression coefficients of the calculations of **Set I**, the ones of **Set II** range from 0.70 to 0.98 depending on the salt concentration. In this case, primary minimum flocculation does not occur. The total number of drops is preserved during the course of the calculations. Higher regression coefficients correspond to higher salinities, where fast irreversible flocculation is more likely to occur. In the case of 300 mM NaCl, the actual curve of $1/n_{agg}$ vs. time for the simulations of **Set II** oscillates (not shown), indicating that total number of aggregates changes erratically at all times. This suggests reversible secondary minimum flocculation with no stable dimers formed [Urbina-Villalba, 2005b, 2006]. The situation is clearly illustrated in Figure 7. The number of single particles slightly decreases, and the number of doublets appears to be constant during the whole simulation. More strictly, the number of monomers and dimers fluctuate around an average value. In fact, the number of monomers still constitutes 93% of the total aggregate population at t = 30 s (Table 2) and the amount of dimers does not surpass 5%. As the salinity increases (NaCl = 600 mM), the interaction potential decreases, and the aggregates resemble more closely the predictions of Eq. (2). This is illustrated in Figure 8 (**Set II** simulations).

Figures 9 and 10 show a comparison between the experimental values of $k_{FC}$, and the ones obtained from the simulations of **Set I** and **II**, respectively. The calculations basically differ in the charge of the surfactant molecule and the variation of its surface concentration at the interface of the drops due to a change in the salt concentration. In the case of **Set I**, the surfactant distribution is consistent with the findings of Gurkov *et al.* [2005] for macroscopic adsorption isotherms. In the case of **Set II** a novel procedure which allows the evaluation of the surface concentration as a function of either the interfacial tension or the surface potential is employed. It is evident that the macroscopic adsorption isotherms (**Set I**) do not reproduce the experimental findings. The resulting theoretical rates are at least four times faster than the ones experimentally observed.

Instead, the simulations of **Set II** show the correct order magnitude differing at most in one unit from the experimental rates. The qualitative form of the theoretical curve is correct, but the simulations show a stronger dependence with the salt concentration than expected. The curves cross each other around 500 mM. Hence, it appears that the repulsive barrier of the interaction potential is overestimated below this threshold, and it is underestimated above it. Still, the qualitative trend is correct. Moreover, the results of Figure 10 correspond to a simulation time of 30 seconds. They are expected to improve slightly at higher times, since it is commonly observed that the slope of $1/n_{agg}$ vs. t progressively decreases as a function of time.

While the qualitative behavior of the colloids can be justified in terms of DLVO, the quantitative evaluation of stabil-ity is not consistent with this theory. It is a common practice to justify a given flocculation rate on the basis of a convenient stability ratio. But in the simulations of **Set II**, *no particle crosses the repulsive barrier* in its (unsuccessful) path towards primary minimum flocculation. If the particles do not reach the primary minimum and the potentials curves converge for separation distances larger than 4 nm (see Fig. 5) the differences in the values of $k_{FC}$ (Fig. 9) can only be related to: a) the effect of the secondary minimum (which moves towards lower separation distances and is more profound as the NaCl concentration increases), and b) to the slope of the potential energy curve in the vicinity of the secondary minimum. This is known as the "hardness" of the potential barrier, namely, the rapid variation of $dV/dr_{ij}$ at distances of separation shorter than the position of the secondary minimum. As additional two-particle simulations demonstrated, two aggregated particles spend only 25% of the time at such distances in the case of 300 mM NaCl, while being 40% of the time in this region for 1000 mM NaCl (In these additional calculations, a small cubic box (L = 5R) was used. It contained two particles at a starting distance of 4 nm. The simulations were run for 2 billion iterations). In any event, the "delay" of the particles in their strive towards primary minimum flocculation cannot be quantified in terms of the formalism proposed by Fuchs [Fuchs, 1936; McGown, 1967], since the drops do not cross the repulsive barrier during the course of the simulation.

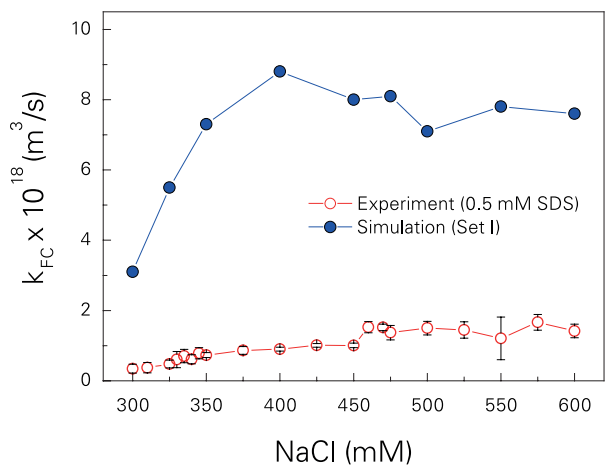

**Figure 9**: Comparison between the experimental results, and the predictions of the simulations corresponding to **Set I** (Adsorption isotherms).

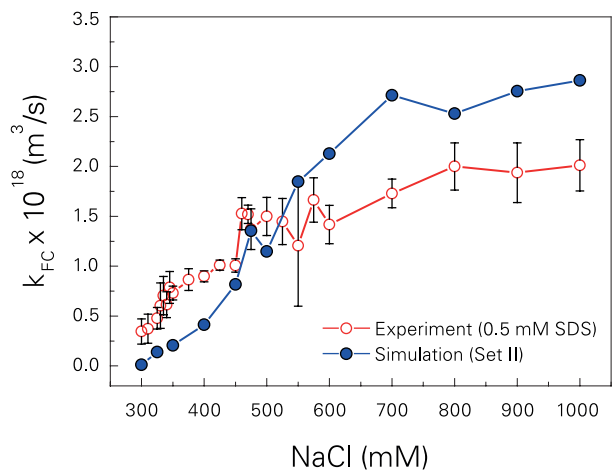

**Figure 10**: Comparison between the experimental results, and the predictions of the simulations corresponding to **Set II** (Electrostatic surface potentials).

The referred inconsistency between DLVO and the present results becomes evident, if the equation of Lozsán *et al.*

[Lozsán, 2006] is used to calculate the magnitude of the potential barriers which are consistent with the experimental behavior found:

$$\ln W = 0.40\left(\frac{\Delta V}{k_B T} - 1\right) \tag{30}$$

Equation (30) was deduced from ESS of hexadecane-in-water emulsions stabilized with nonyl phenol ethoxylated (NPE$_m$, $6 \leq m \leq 17$) surfactants. The equation is very similar to the one previously deduced by Prieve and Ruckenstein [Prieve, 1977] from the numeric evaluation of the analytical formula of the stability ratio for the case of rigid particles. An average value of the fast flocculation rate can be deduced from **Set I** simulations: $k_f^{fast} = 7.75 \times 10^{-18}$ m$^3$/s (Fig. 9). This rate can be used to calculate the stability ratio required to reproduce the experimental data. Following this procedure, stability ratios W (= $k_f^{fast} / k_f^{slow}$, where: $k_f^{slow} = k_f^{exp}$) between $3.9 \leq W \leq 22.4$ result. Hence, according to Eq. 30, repulsive barriers between $3.7 \leq \Delta V/k_B T \leq 7.3$ are required to reproduce the experiments. However, as shown by the calculations of **Set I** (see Fig. 3), these barriers are too small to prevent fast flocculation and coalescence. On the other hand, it is evident from the magnitude of the barriers shown in Fig. 4 that if one calculates the stability ratio of the interaction potentials corresponding to the simulations of **Set II** (Fig. 4), huge numbers result. For example, W $\approx 10^{34}$ for NaCl = 400 mM. It is clear that it is not possible to reconcile the values predicted by DLVO with the experimental data whenever the particles do not surmount the repulsive barrier.

The effect of the secondary minimum on the rate of dimer formation ($k_{11}$) has been formally considered by several authors (see Refs. [Sonntag, 1987; Behrens and Borkovec, 2000; Urbina-Villalba, 2005b; Ohshima, 2013] and references therein). In general, a shallow secondary minimum is expected to decrease the magnitude of an effective flocculation rate due to the redissolution of the dimers. Recently, Ohshima deduced simple analytical equations for the temporal variation of monomers and dimers in the early stage aggregation of a suspension. The formalism considers reversible flocculation of the monomers (M) in the secondary minimum and irreversible flocculation of the dimers in the primary minimum (P):

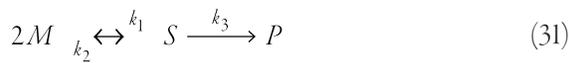

$$2M \underset{k_2}{\overset{k_1}{\longleftrightarrow}} S \xrightarrow{k_3} P \qquad (31)$$

The analytical expressions of $n_1(t)$, $n_{2P}(t)$, $n_{2S}(t)$ are very close to the exact numerical calculations reported by Behrens and Borkovec [2000]. They depend on the sum and the difference of two average rates:

$$K_a = \frac{k_2 + k_3}{2} + k_1 n_0 \qquad (32)$$

$$K_b = \left( K_a^2 - 2\, k_1\, k_3\, n_0 \right)^{1/2} \qquad (32)$$

These constants appear in the equations of $n_1(t)$, $n_{2P}(t)$, $n_{2S}(t)$ as exponentially decay rates: $(K_a + K_b)$ and $(K_a - K_b)$, which allow to define two relaxation times, one corresponding to a slow aggregation and one related with fast aggregation. As a result, the total number of dimers $n_2(t)$ = $n_{2P}(t) + n_{2S}(t)$ shows an abrupt initial increase, and a much slower (almost linear) one. Using values of $n_0 = 10^{15}$ m$^{-3}$, $k_1$ = 1.1 x $10^{-17}$ s$^{-1}$, $k_2 = 6.6$ x $10^{-2}$ s$^{-1}$, $k_1 = 3.2$ x $10^{-4}$ s$^{-1}$, relaxation times of $\tau_1 = 1/(K_a + K_b) = 11$ s, and $\tau_2 = 1/(K_a + K_b) = 208$ min, were obtained.

The underlying rate constants of the process described in Eq. (31): $k_1$, $k_2$, and $k_3$, are expressed as integrals of the pair interaction energy which depend on: the position of the secondary minimum ($r_S$), the position of the energy barrier ($r_B$), the distance at which two particles are considered to form a dimer, ($r_D$), and the minimum distance of separation ($2R$). Notice that the force between two particles i and j is equal to $-dV/dr_{ij}$ (Eq. (7)). Thus, the so-called "hardness" of the potential corresponds to the value of this derivative between $r_S$ and $r_B$. Moreover, if primary minimum flocculation does not occur, $k_3 = 0$. In this case, $K_a = K_b = k_2/2 + k_1 n_0$.

The referred analysis could probably be applied to the systems of lowest salinities (NaCl < 350 mM). However, as it is shown in Table 2 and Fig. 8 for the calculations of **Set II**, much larger clusters are formed during the thirty seconds of evolution of the emulsions. But then, if the repulsive barrier does not modulate the value of the rate constant, why does the flocculation rate of the actual systems progressively increase as a function of the ionic strength? In this regard, the findings of Ohshima, Behrens and Borkovec can be very insightful. We believe that the delay in the average flocculation rate of the system ($k_{FC}$) does not come from a delay in the production of the dimers, which will afterwards turn into trimers and so on, following a Smoluchowskian scheme of aggregation. This is a possibility that will certainly decrease the average aggregation rate under the constant kernel approximation. However, in our view, the "delay" is rather related to the absence of stable aggregates of any size [Urbina-Villalba, 2005b]. If each monomer experiences reversible flocculation with their neighbors *within an aggregate*, the average number of particles in the aggregates will change as a function of time. As a result, aggregates will be partially disassembled and re-assembled continuously, delaying the decrease of the total number of aggregates of any size as a function of time. As the salinity increases, the secondary minimum is deeper and the time spent by the particles in the vicinity of the secondary minimum increases. Hence, flocs are more stable and the average aggregation rate increases. This behavior is consistent with the curves of $n_k$ vs. t usually produced by the simulations when a shallow secondary minimum precedes a large repulsive barrier: they appear to be the result of a superposition of a rapidly fluctuating oscillation and a monotonic decrease.

## 6. CONCLUSIONS

The comparison between the experimental values of the flocculation rate of hexadecane-in-water nano-emulsions measured in our laboratory with the ones theoretically evaluated by means of Emulsion Stability Simulations of the same systems, suggests that ESS are a reliable tool for the theoretical study of these dispersions. The fact that the flocculation rates predicted by the simulations of **Set II** and the ones experimentally measured reasonably coincide, reaffirm the validity of both techniques. In particular, it is an indication that Eq. (1) is trustworthy, and that the resulting values of $k_{FC}$ obtained from the adjustment of Eq. (1) to the experimental turbidity of the emulsions are relevant, actually reflecting the temporal evolution of these systems.

As shown in this report, the surface charge of nano-drops cannot be deduced using the surface excess of "macroscopic" adsorption isotherms. A purposely designed procedure which takes into account the particular variation of the surface charge of the emulsion drops as a function of the salt concentration is needed [Urbina-Villalba, 2013].

In the case of solid particles it is not possible to distinguish experimentally secondary minimum aggregation from primary minimum flocculation. However, the coalescence of non-deformable drops only follows after primary minimum aggregation, and therefore allows such a distinction. According to our simulations, hexadecane drops stabilized with 0.5 NaCl do not cross the potential barrier of their interaction potential, and therefore, do not coalesce. Hence, the reasonable agreement between the values of $k_{FC}$ experimentally found and the ones predicted by the simulations of **Set II**, strongly indicates that the experimental behavior cannot be justified in terms of the passage (diffusion of the particles) *over* the repulsive barrier [Fuchs, 1936; McGown, 1967]. In the present case, the change of $k_{FC}$ as a function of the ionic strength is rather connected with the reversible aggregation of the clusters of any size. The stability of the aggregates towards disintegration depends on the free energy of interaction with their neighbors. As a result, the differences in the flocculation rates observed are probably caused by the distinct times of residence of the drops in the secondary minimum of the pair potential. These times depend on the depth of the secondary minimum and more importantly, on the hardness of the potential barrier (how fast it grows at very short distances of separation).

## 7. APPENDIX

More elaborate corrections of the adsorption isotherms are possible when equations (15) or (19) fail. This occurs in general at very low ionic strength (low salt concentration and/or low surfactant concentration) or above the CMC. We found convenient to introduce two auxiliary equations for this purpose, based on the logarithmic dependence of the tension on the surfactant concentration suggested by the Gibbs adsorption isotherm. For a surfactant concentration $C_s \geq 10^{-4}$ M, the tension can be reasonably approximated by:

$$\gamma = \gamma_0 + \left[ \frac{\gamma_0 - \gamma_c}{\ln(C_{s,min}) - \ln(CMC)} \right] (\ln(C_s) - \ln(C_{s,min}))$$

(A.1)

Where $C_{s,min} = 10^{-6}$ M. For $C_s < 10^{-4}$ M the values of $\gamma_c$ and CMC are substituted by the value of $\gamma(C_s = 10^{-4}$ M$)$ and $10^{-4}$ M, respectively. In the presence of salt $\gamma(C_s = 10^{-4}$ M$)$ is calculated using Eq. (15). Otherwise, the expression of Rehfeld is used:

$$\gamma = -1.62 (\ln C_s)^2 - 31.5 (\ln C_s) - 102.3 \quad (\text{mN/m})$$

(A.2)

In turn, the interfacial area of the surfactant ($A_s$) can be approximated by Eq. (A.3):

$$A_s = A_{s,max} + \left[ \frac{A_{s,max} - A_\infty}{\ln(C_{s,min}) - \ln(CMC)} \right] (\ln(C_s) - \ln(C_{s,min}))$$

(A.3)

Where $A_{s,max} = 100 * A_\infty$.

The procedure described by Eqs. (A.1) – (A.3) produces the isotherms depicted in Figure 11.

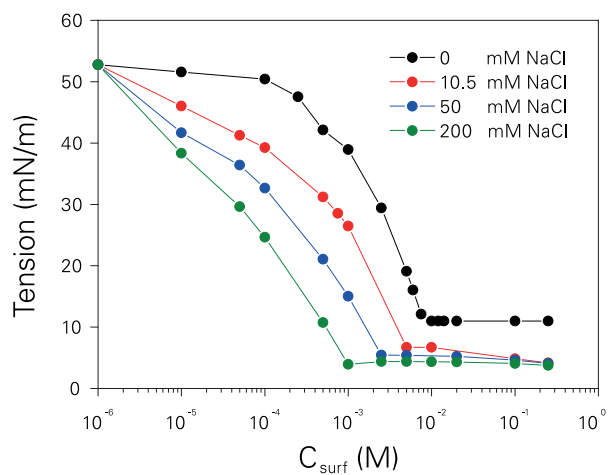

**Figure 11:** Approximate adsorption isotherms deduced from implementing the corrections of Eqs. (A.1) to (A.3).

It was evident from the calculation of the interaction potentials which correspond to the isotherms of Figs. 2 and 11, that the potential barriers of the latter are smaller (not shown). Hence, these isotherms cannot improve the results of the simulations of **Set I** despite their higher accuracy; since they would necessarily lead to either similar or faster flocculation rates.